\documentclass[12pt]{article}
\usepackage{float}
\usepackage{braket}
\usepackage[date=year,style=numeric,giveninits=true,sorting=none,maxnames=20]{biblatex}
\AtEveryBibitem{\clearlist{language} \clearlist{location} }
\renewbibmacro{in:}{}
\DeclareFieldFormat[article, inbook,proceedings]{title}{#1}
\usepackage[dvipdfm,margin=1.1in]{geometry}
\usepackage{amsfonts,mathtools,amsthm,bm,amssymb}
\usepackage{subfiles}
\usepackage{subcaption}
\usepackage[dvipdfmx]{graphicx}
\usepackage{authblk}
\usepackage[dvipdfmx]{color}
\usepackage{hyperref}
\hypersetup{
setpagesize=false,
 colorlinks=true,%
 linkcolor=magenta,
 citecolor=blue,
}
\newtheorem{theorem}{Theorem}[section]

\newtheorem{Corollary}[theorem]{Corollary}
\newtheorem{Lemma}[theorem]{Lemma}
\newtheorem{Proposition}[theorem]{Proposition}

\title{\bf Return probability of quantum and correlated random walks}
\author[1]{Chusei Kiumi*}
\affil[1]{\footnotesize Graduate School of Science and Engineering, Yokohama National University, 
\protect\\ 
Hodogaya, Yokohama, 240-8501, Japan,
\protect\\ 
E-mail: kiumi-chusei-bf@ynu.jp}

\author[2]{Norio Konno}
\affil[2]{\footnotesize Department of Applied Mathematics, Faculty of Engineering,Yokohama National University,
\protect\\ 
Hodogaya, Yokohama, 240-8501, Japan
\protect\\
E-mail: konno-norio-bt@ynu.ac.jp}

\author[3]{Shunya Tamura}
\affil[3]{Graduate School of Science and Engineering, Yokohama National University, 
\protect\\ 
Hodogaya, Yokohama, 240-8501, Japan
\protect\\ 
E-mail: tamura-shunya-kj@ynu.jp} 

\date{\empty}

\begin{document}
\maketitle
\vspace{-1.2cm}
\begin{abstract}
The analysis of the return probability is one of the most essential and fundamental topics in the study of classical random walks. In this paper, we study the return probability of quantum and correlated random walks in the one-dimensional integer lattice by the path counting method. We show that the return probability of both quantum and correlated random walks can be expressed in terms of the Legendre polynomial. Moreover, the generating function of the return probability can be written in terms of elliptic integrals of the first and second kinds for the quantum walk.
\end{abstract}
\section{Introduction}
The classical random walk is one of the most important and widely used models in the scientific research \cite{kac1947random, metzler2000random,ceperley1986quantum}. 
Also, a more general model called correlated random walk \cite{bohm2000correlated} has been actively studied for modeling more complex motions \cite{kareiva1983analyzing,codling2008random}. The movement of the correlated random walk depends on the motion of the previous step. Furthermore, since the early 2000s, research on the quantum-mechanical analogue of the random walk has been attracting much attention. The model is called quantum walk and plays essential roles in various fields. For a comprehensive review, see \cite{venegas2012quantum}. As an interesting application, quantum walks have been extensively used for quantum field theory, for example, see \cite{di2013quantum, arnault2016quantum1,arnault2016quantum2,manighalam2021continuous}.
Quantum walks have many characteristic properties that are not present in classical random walks, such as ballistic spreading \cite{konno2002quantum} and localization \cite{inui2005one}, which further expand the potential for applications. For this reason, it is significant to compare the fundamental properties of classical and quantum walks. 

In this paper, we focus on the return probability of discrete-time quantum and correlated random walks in the one-dimensional integer lattice, where the walker starts from the origin. The return probability has been actively studied since it is one of the most important and fundamental research topics in classical random walks. In the case of quantum walks, there is a deep connection with localization property, which is vital for applications in different studies such as quantum search algorithms \cite{Ambainis2005-ha,Childs2004-xa,Shenvi2003-jw} and topological insulators \cite{Kitagawa2010-su,Endo2015-db}. For the relation between return probability and localization, see \cite{cantero2012one} for detail. A fundamental study on the return probability of the random walk was carried out by G. Pólya \cite{Polya1921}. It is proved that the generating function of the return probability of the two-dimensional random walk and the return probability of the three-dimensional random walk can be written in terms of the elliptic integral of the first kind (see also \cite{spitzer2001principles}). A similar expression of the generating function of the return probability was given in \cite{konno2010quantum} for the one-dimensional quantum walk. However, the result was restricted to a specific model called the Hadamard walk, and its initial state was also specified. In this paper, we extend this result and prove that the expression can be written with elliptic integrals of both the first and second kinds for the general time evolution and initial state. Moreover, we show that the return probability of both quantum and correlated random walks can be written by the Legendre Polynomial. Similar to the elliptic integrals, the Legendre polynomial is also a well-studied special function, useful for a variety of analyses \cite{andrews1999special}. Particularly in this paper, the return probability is expressed in terms of the Legendre polynomial, allowing for the further analysis of the generating function and characterization of the return probability. For more about the return probability of the quantum walk, see \cite{vstefavnak2008recurrence1,vstefavnak2008recurrence2,vstefavnak2009recurrence,xu2010discrete,Ide2011return,machida2015limit}.

The rest of this paper is organized as follows. In Section \ref{sec:qw}, we focus on the analysis of the quantum walk. After giving the definition, we show that the return probability can be represented with the Legendre polynomial in Proposition \ref{prop:qw_return}, and its generating function can be written with elliptic integrals in Proposition \ref{prop:qw_generating}. Subsequently, Section \ref{sec:crw} is devoted to the analysis of the correlated random walk. Proposition \ref{prop:crw_return} proves that the return probability of the correlated random walk can also be represented with the Legendre polynomial, and Proposition \ref{prop:crw_generating} provides a result for its generating function. Finally, the conclusion and further discussion are presented in Section \ref{sec:summary}.

\section{Quantum walk}
\label{sec:qw}
\subsection{Definition}
In this section, we consider a discrete-time one-dimensional quantum walk. As for a detailed information of the definition, see \cite{venegas2012quantum,Portugal} for example. First, we define the coin matrix $U$ by $2\times 2$ unitary matrix given as
\begin{equation*}
U=\left[\begin{array}{ l l }
a & b\\
c & d
\end{array}\right] \ \ (a,b,c,d\in \mathbb{C} ),
\end{equation*}
where $\mathbb{C}$ denotes a set of complex numbers. The quantum walk has a degree of freedom called chirality, which takes the value of left or right, meaning the direction of the walker's motion. At each time step, a walker with left chirality will move one unit to the left, and a walker with right chirality will move one unit to the right. We consider
\begin{equation*}
|L\rangle =\left[\begin{array}{ l }
1\\
0
\end{array}\right] ,\ \ |R\rangle =\left[\begin{array}{ l }
0\\
1
\end{array}\right] ,
\end{equation*}
where $L$ and $R$ represent the left chirality and right chirality states, respectively. We divide $U$ into the following two matrices to define the dynamics of the model.
\begin{equation*}
P=\left[\begin{array}{ l l }
a & b\\
0 & 0
\end{array}\right] ,\ \ Q=\left[\begin{array}{ l l }
0 & 0\\
c & d
\end{array}\right] .\ 
\end{equation*}
Here, $U=P+Q$ and $P$ represents that the walker moves one unit to the left, and $Q$ represents that the walker moves one unit to the right. Then, we let $\Xi _{n}^{( QW)}( l,m)$ denotes the sum of all paths starting from the origin, where $l$ is a number of steps to the left and $\displaystyle m$ is a number of steps to the right. Note that $n=l+m$ holds. For example, when $n=3$, we have the followings:
\begin{equation*}
\begin{aligned}
 & \Xi _{3} (0,3)=Q^{3} ,\ \ \Xi _{3} (1,2)=Q^{2} P+QPQ+PQ^{2} ,\\
 & \Xi _{3} (2,1)=P^{2} Q+PQP+QP^{2} ,\ \ \Xi _{3} (3,0)=P^{3} .
\end{aligned}
\end{equation*}
Next, we set $S_{n}^{( QW)}$ as a position of the walker at time $n$, which starts from the origin with the initial state $\varphi $:
\begin{equation*}
P\left( S_{n}^{( QW)} =x\right) =\left\Vert \Xi _{n}^{( QW)} (l ,m)\varphi \right\Vert ^{2} ,
\end{equation*}
where $n=l+m,\ x=-l+m$ and
\begin{equation*}
\varphi =\varphi _{1} |L\rangle +\varphi _{2} |R\rangle \in \mathbb{C}^{2} ,\ |\varphi _{1} |^{2} +| \varphi _{2}| ^{2} =1.
\end{equation*}
We define the return probability $r_{n}^{( QW)} (0)$ as the probability that the walker returns to the origin at time $n$.
\begin{equation*}
r_{n}^{( QW)} (0)=P( S_{n} =0) .
\end{equation*}
The previous study \cite{konno2010quantum} gave a return probability $r_{n}^{( QW)} (0)$ and the generating function of $r_{n}^{( QW)} (0)$ for the Hadamard walk whose coin matrix is defined by the Hadamard matrix $H$:
\begin{equation*}
U=H=\frac{1}{\sqrt{2}}\left[\begin{array}{ c c }
1 & 1\\
1 & -1
\end{array}\right] .
\end{equation*}
Also, the initial state was restricted to
\begin{equation*}
\varphi =\frac{1}{\sqrt{2}} |L\rangle +\frac{i}{\sqrt{2}} |R\rangle .
\end{equation*}
However, we extend the results for the general coin matrix and initial state $\varphi $. 

\subsection{Return probability of the quantum walk}
To derive the return probability $r_{2n}^{( QW)} (0)$, first we consider $\Xi _{2n}^{( QW)}( n,n)$. The following lemma is given in the previous study \cite{konno2002quantum}.
\begin{Lemma}
\label{lemma:path}
\begin{equation*}
\Xi _{2n} (n,n)=a^{n} d^{n}\sum _{\gamma =1}^{n}\left(\frac{bc}{ad}\right)^{\gamma }\left(\begin{array}{ l }
n-1\\
\gamma -1
\end{array}\right)^{2}\left(\frac{n-\gamma }{a\gamma } P+\frac{n-\gamma }{d\gamma } Q+\frac{1}{c} R+\frac{1}{b} S\right) .
\end{equation*}
Here,
\begin{equation*}
R=\left[\begin{array}{ l l }
c & d\\
0 & 0
\end{array}\right] ,\ \ S=\left[\begin{array}{ l l }
0 & 0\\
a & b
\end{array}\right] .
\end{equation*}
\end{Lemma}
Next, we give the general expression of a unitary matrix for the coin matrix as follows:
\begin{equation*}
U=\left[\begin{array}{ l l }
a & b\\
c & d
\end{array}\right] =e^{i\theta }\left[\begin{array}{ c c }
\alpha  & \beta \\
-\overline{\beta } & \overline{\alpha }
\end{array}\right] ,
\end{equation*}
where $\theta \in [ 0,2\pi ) ,\alpha ,\beta \in \mathbb{C}$ and $|\alpha |^{2} +|\beta |^{2} =1$. To exclude obvious cases, we assume $\alpha, \beta\neq 0$. $r_{2n}^{( QW)} (0)$ can be expressed by the Legendre polynomial. As for the special function, see \cite{andrews1999special}.
\begin{Proposition}
\label{prop:qw_return}
\begin{equation*}
r_{2n}^{( QW)} (0)=\frac{\{P_{n-1} (k)\}^{2} -2kP_{n} (k)P_{n-1} (k)+\{P_{n} (k)\}^{2}}{2(k+1)} ,\ r_{2n-1}^{( QW)} (0)=0
\end{equation*}
for $n\geq 1$ and $r_{0}^{( QW)}( 0) =1$. Here, $k=2|\alpha |^{2} -1$ and $P_n(x)$ denotes the Legendre polynomial.
\begin{proof}
The proof will be stated in Appendix \ref{appendix1}.
\end{proof}
\end{Proposition}
The result shows that the return probability does not depend on the initial state $\varphi $, and it only depends on $|\alpha |$. Putting $|\alpha |=\frac{1}{\sqrt{2}}$, we get the result from the previous study \cite{konno2010quantum} as a corollary. \ 
\begin{Corollary}
The return probability of the Hadamard walk $\displaystyle r_{2n}^{( H)}( 0)$ becomes $r_{0}^{( H)}( 0) =1$,
\begin{equation*}
r_{2n}^{( H)} (0)=\frac{1}{2}\left[\{P_{n-1} (0)\}^{2} +\{P_{n} (0)\}^{2}\right] ,\ r_{2n-1}^{( H)}( 0) =0
\end{equation*}
for $n\geq 1$. Since
\begin{equation*}
P_{2n+1} (0)=0,\ P_{2n} (0)=\frac{1}{2^{2n}}\left(\begin{array}{ c }
2n\\
n
\end{array}\right)
\end{equation*}
hold, we get
\begin{equation*}
r_{4m}^{( H)} (0)=r_{4m+2}^{( H)} (0)=\frac{1}{2}\{P_{2m} (0)\}^{2} =\frac{1}{2^{4m+1}}\left(\begin{array}{ c }
2m\\
m
\end{array}\right)^{2} \ \ (m\geq 1).
\end{equation*}
\end{Corollary}
Nextly, we consider the generating function of the return probability.
\begin{Proposition}
\label{prop:qw_generating}
For $k=2|\alpha |^{2} -1$,

\begin{align*}
 & \sum _{n=0}^{\infty } r_{n}^{( QW)} (0)z^{n} =\frac{1}{\pi (k+1)}\left(\left( 1+z^{2}\right)\mathcal{K}\left( k,z^{2}\right) -2k^{2}\int _{0}^{z^{2}}\frac{\mathcal{E}( k,w)}{1-w} dw-\frac{\pi}{2}  \right) +1,
\end{align*}
where 
\begin{equation*}
\mathcal{K}( x,z) =\frac{K\left(\sqrt{\frac{4z\left( 1-x^{2}\right)}{1-2z\left( 2x^{2} -1\right) +z^{2}}}\right)}{\sqrt{1-2z\left( 2x^{2} -1\right) +z^{2}}} ,\qquad \mathcal{E}( x,z) =\frac{E\left(\sqrt{\frac{4z\left( 1-x^{2}\right)}{1-2z\left( 2x^{2} -1\right) +z^{2}}}\right)}{\sqrt{1-2z\left( 2x^{2} -1\right) +z^{2}}}.
\end{equation*}
Here, $K$ and $E$ are elliptic integrals of the first and second kind, respectively. They are defined by 
\begin{align*}
K( m) & =\int\nolimits _{0}^{\frac{\pi }{2}}\frac{d\theta }{\sqrt{1-m^{2}\sin^{2} \theta }} =\int _{0}^{1}\frac{dx}{\sqrt{\left( 1-x^{2}\right)\left( 1-m^{2} x^{2}\right)}} \quad ( 0\leq m< 1) ,\ \\
E( m) & =\int\nolimits _{0}^{\frac{\pi }{2}}\sqrt{1-m^{2}\sin^{2} \theta } \ d\theta =\int _{0}^{1}\sqrt{\frac{1-m^{2} x^{2}}{1-x^{2}}} \quad ( 0\leq m< 1) .
\end{align*}
\begin{proof}
 It is known that the generating function of the product of two Legendre polynomials can be expressed as follows (see \cite{maximon1956generating}):
\begin{equation*}
\sum _{n=0}^{\infty } P_{n}(\cos \theta _{1}) P_{n}(\cos \theta _{2}) z^{n} =\frac{_{2} F_{1}\left(\frac{1}{2} ,\frac{1}{2} ;1;\frac{4z\sin \theta _{1}\sin \theta _{2}}{1-2z\cos (\theta _{1} +\theta _{2} )+z^{2}}\right)}{\sqrt{1-2z\cos (\theta _{1} +\theta _{2} )+z^{2}}} .
\end{equation*}
Thus, we get
\begin{align}
\label{eq:square_legendre}
\sum _{n=1}^{\infty }\{P_{n}( x)\}^{2} z^{n} =\frac{2}{\pi }\mathcal{K}( x,z) -1
\end{align}
and
\begin{equation*}
\sum _{n=1}^{\infty }\{P_{n-1}( x)\}^{2} z^{n} =\frac{2z}{\pi }\mathcal{K}( x,z)
\end{equation*}
for $|x|\leq 1$. Next, using the following relation of the Legendre polynomial,
\begin{equation*}
P_{n-1} (x)=xP_{n} (x)-\frac{x^{2} -1}{n}\frac{d}{dx} P_{n} (x),
\end{equation*}
we have
\begin{align}
\sum _{n=1}^{\infty } z^{n} P_{n} (x)P_{n-1} (x) & =\sum _{n=1}^{\infty } z^{n} P_{n} (x)\left( xP_{n} (x)-\frac{x^{2} -1}{n}\frac{d}{dx} P_{n} (x)\right)\nonumber\\
 & =\sum _{n=1}^{\infty } z^{n}\left[ x\{P_{n} (x)\}^{2} -\frac{x^{2} -1}{n} P_{n} (x)\frac{d}{dx} P_{n} (x)\right]\nonumber \\
 & =\sum _{n=1}^{\infty } z^{n} x\{P_{n} (x)\}^{2} -\left( x^{2} -1\right)\sum _{n=1}^{\infty }\frac{z^{n}}{n} P_{n} (x)\frac{d}{dx} P_{n} (x)\nonumber\\
 \label{eq:product_mid}
 & =x\sum _{n=1}^{\infty }\{P_{n} (x)\}^{2} z^{n} -\frac{x^{2} -1}{2}\int _{0}^{z} w^{-1}\frac{d}{dx}\sum _{n=1}^{\infty }\{P_{n} (x)\}^{2} w^{n} dw.
\end{align}
Note that 
\begin{equation*}
\begin{aligned}
 & \frac{d}{dz}\mathcal{K}( x,z) =\frac{( 1+z)\mathcal{E}( x,z) -( 1-z)\mathcal{K}( x,z)}{2z( 1-z)} ,\\
 & \frac{d}{dx}\mathcal{K}( x,z) =x\frac{\mathcal{E}( x,z) -\mathcal{K}( x,z)}{x^{2} -1} .
\end{aligned}
\end{equation*}
By (\ref{eq:square_legendre}), we can rewrite (\ref{eq:product_mid}) with elliptic integrals 
\begin{equation*}
\sum _{n=1}^{\infty } z^{n} P_{n} (x)P_{n-1} (x)=\frac{2x}{\pi }\mathcal{K}( x,z) -x-\frac{x}{\pi }\int _{0}^{z}\frac{\mathcal{E}( x,w) -\mathcal{K}( x,w)}{w} dw.
\end{equation*}
By differentiating with respect to $z$ for both sides, we get
\begin{equation*}
\sum _{n=1}^{\infty } nz^{n-1} P_{n} (x)P_{n-1} (x)=\frac{2x\mathcal{E}( x,z)}{\pi ( 1-z)} .
\end{equation*}
Therefore, we have a simple expression as follows:
\begin{align*}
\sum _{n=1}^{\infty } z^{n} P_{n} (x)P_{n-1} (x) & =\int _{0}^{z}\sum _{n=1}^{\infty } nw^{n-1} P_{n} (x)P_{n-1} (x)dw\\
 & =\frac{2x}{\pi }\int _{0}^{z}\frac{\mathcal{E}( x,w)}{1-w} dw.
\end{align*}
It follows from these discussions that 
\begin{equation*}
\begin{aligned}
\sum _{n=0}^{\infty } r_{n}^{( QW)} (0)z^{n} & =\left(\sum _{n=1}^{\infty } r_{2n}^{( QW)} (0)z^{2n}\right) +1\\
 & =\left(\frac{1}{2(k+1)}\sum _{n=1}^{\infty }\left(\{P_{n-1} (k)\}^{2} -2kP_{n} (k)P_{n-1} (k)+\{P_{n} (k)\}^{2}\right) z^{2n}\right) +1\\
 & =\frac{1}{\pi (k+1)}\left\{\left( 1+z^{2}\right)\mathcal{K}\left( k,z^{2}\right) -2k^{2}\int _{0}^{z^{2}}\frac{\mathcal{E}( k,w)}{1-w} dw-\frac{\pi }{2}\right\} +1.
\end{aligned}
\end{equation*}
\end{proof}
\end{Proposition}
Setting $|\alpha |=\frac{1}{\sqrt{2}}$, \ we obtain the result from previous study \cite{konno2010quantum} as a corollary. \ 
\begin{Corollary}
The generating function of the return probability for the Hadamard walk becomes
\begin{align*}
\sum _{n=0}^{\infty } r_{n}^{( H)}( 0) z^{n} & =\frac{1+z^{2}}{\pi } K\left( z^{2}\right) +\frac{1}{2}.
\end{align*}
\begin{proof}
Putting $k=0$, we have
\begin{equation*}
\begin{aligned}
\sum _{n=0}^{\infty } r_{n} (0)z^{n} & =\frac{1}{\pi }\left\{\left( 1+z^{2}\right)\mathcal{K}\left( 0,z^{2}\right) -\frac{\pi }{2}\right\} +1\\
 & =\frac{K\left(\frac{2z}{1+z^{2}}\right)}{\pi } +\frac{1}{2} .
\end{aligned}
\end{equation*}
We can use the following relation from \cite{gradshteyn2014table}:
\begin{equation*}
K\left(\frac{2\sqrt{| t| }}{1+t}\right) =(1+t)K(t)
\end{equation*}
for $| t| < 1$. Thus, we see
\begin{equation*}
\sum _{n=0}^{\infty } r_{n}^{( H)} (0)z^{n} =\frac{1+z^{2}}{\pi } K\left( z^{2}\right) +\frac{1}{2} .
\end{equation*}
\end{proof}
\end{Corollary}

As a remark, we introduce the return probability of the two-dimensional random walk obtained by Pólya \cite{Polya1921} (see also Spitzer \cite{spitzer2001principles}) as below:
\begin{equation*}
r_{2n}^{(RW,2)} (0)=\frac{1}{4^{2n}}\left(\begin{array}{ c }
2n\\
n
\end{array}\right)^{2},\quad \ p_{2n+1}^{(RW,2)} (0)=0\ \ (n\geq 0).
\end{equation*}
Therefore, we have an expression with the elliptic integral of the first kind.
\begin{equation*}
\sum _{n=0}^{\infty } r_{n}^{(RW,2)} (0)z^{n} =\frac{2}{\pi } K(z).
\end{equation*}
Also, the probability that the three-dimensional random walk returns to its starting point, $F=1-G^{-1} ,$ is given by \cite{spitzer2001principles} as
\begin{equation*}
G=\frac{1}{\pi ^{2}}\int _{-\pi }^{\pi } K\left(\frac{2}{3-\cos \theta }\right) d\theta .
\end{equation*}
\section{Correlated random walks}
\label{sec:crw}
\subsection{Definition}
We consider a one-dimensional correlated random walk where the probability of moving to the next step depends on the previous step. The evolution is defined as follows:
\begin{align*}
 & P\left(\text{particle moves one unit to the left}\right)\\
 & \ \ \ \ \ \ \ =\begin{cases}
p, & \text{if the previous step was to the left,}\\
1-q, & \text{if the previous step was to the right,}
\end{cases}
\end{align*}
and
\begin{align*}
 & P\left(\text{particle moves one unit to the right}\right)\\
 & \ \ \ \ \ \ \ \ =\begin{cases}
1-p, & \text{if the previous step was to the left,}\\
q, & \text{if the previous step was to the right. }
\end{cases}
\end{align*}
Here, if $p=q,$ the walker moves one unit in the same direction with probability $p$, or the walker moves one unit in the opposite direction with probability $1-p$. In the case of $p=1-q,$ the walk becomes uncorrelated with the past time. Thus, we see that correlated random walks include random walks as special cases. Furthermore, when $p=q=1/2$, \ the walk is equivalent to the well-known symmetric (non-correlated) random walk, i.e., the particle moves at each step either one unit to the left with probability $1/2$, or one unit to the right with probability $1/2$.

Next, to make the correspondence with quantum walks easier to understand, we define the time evolution of correlated random walks by a $2\times 2$ transition matrix $A$.
\begin{equation*}
A=\left[\begin{array}{ c c }
a & b\\
c & d
\end{array}\right] ,\ 
\end{equation*}
where $a,b,c,d\in [0,1]$ and $a+c=b+d=1$. To exclude the obvious case, we assume $0< a,d< 1$ henceforward. We divide $A$ into the following two matrices to define the dynamics of the model.
\begin{equation*}
\hat{P} =\left[\begin{array}{ l l }
a & b\\
0 & 0
\end{array}\right] ,\ \ \hat{Q} =\left[\begin{array}{ l l }
0 & 0\\
c & d
\end{array}\right] .\ 
\end{equation*}
Here, $A=\hat{P} +\hat{Q}$, and $\hat{P}$ represents that the walker moves one unit to the left, and $\hat{Q}$ represents that the walker moves one unit to the right. Let $\Xi _{n}^{( CRW)} (l,m)$ denote the sum of all paths starting from the origin, where $l$ is a number of steps to the left and $\displaystyle m$ is a number of steps to the right. \ 

Also, let $S_{n}^{( CRW)}$ be the location of the walker at time $n$ starting from the origin with the initial state $\hat{\varphi }$:
\begin{equation*}
P(S_{n}^{( CRW)} =x)=\| \Xi _{n}^{( CRW)} (l ,m)\hat{\varphi } \| _{1}
\end{equation*}
with $n=l+m\ and\ x=-l+m$. Here, $\hat{\varphi } = [\hat{\varphi }_{1} \ \hat{\varphi }_{2} ]^{T}\in \mathbb{R}^{2}$ satisfies $\| \hat{\varphi } \| _{1} =\hat{\varphi }_{1} +\hat{\varphi }_{2} =1$ and $\hat{\varphi }_{1} ,\hat{\varphi }_{2} \geq 0$, where $T$ denotes the transpose operator.
\subsection{Return probability of the correlated random walk}

\begin{Proposition}
\label{prop:crw_return}
\ \\
Case $\Delta _{-} \neq 0:$

Let $\Delta _{\pm } =ad\pm bc$ and $k_{\pm } =ac\hat{\varphi }_{1} +bd\hat{\varphi }_{2} \pm ad$. Then we have 
\begin{equation*}
r_{2n}^{( CRW)}( 0) =\frac{\Delta _{-}^{n}}{2ad}\left( k_{-} P_{n-1}\left(\frac{\Delta _{+}}{\Delta _{-}}\right) +k_{+} P_{n}\left(\frac{\Delta _{+}}{\Delta _{-}}\right)\right) .
\end{equation*}
Note that if $a=d$, the return probability does not depend on the initial state.\\
Case $\Delta _{-} =0:$

The walk becomes random walk. We let $\displaystyle a=b=p,c=d=q=1-p$ and the return probability becomes
\begin{align*}
 & r_{2n}^{( RW)}( 0) =( pq)^{n}\binom{2n}{n} .
\end{align*}
\begin{proof}
The proof will be stated in Appendix \ref{appendix2}.
\end{proof}
\end{Proposition}
\begin{Proposition}
\label{prop:crw_generating}
\ \\ 
Case $\Delta _{-} \neq 0:$

Let $\Delta _{\pm } =ad\pm bc$ and $k_{\pm } =ac\hat{\varphi }_{1} +bd\hat{\varphi }_{2} \pm ad$. Then we get 
\begin{align*}
\sum _{n=0}^{\infty } r_{n}^{( CRW)}( 0) z^{n} & =\frac{1}{2ad}\left(\frac{\Delta _{-} k_{-} z^{2} +k_{+}}{\sqrt{\Delta _{-}^{2} z^{4} -2\Delta _{+} z^{2} +1}} -k_{+}\right) +1.\ 
\end{align*}
Case $\Delta _{-} =0$ (random walk)$:$ we see
\begin{equation*}
\sum _{n=0}^{\infty } r_{n}^{(RW)} (0)z^{n} =\frac{1}{\sqrt{1-z^{2}}} .
\end{equation*}
\begin{proof}
The result is well-known for the case $\Delta_-=0$ (see \cite{spitzer2001principles}, for example), thus we assume $\Delta_-\neq0.$
The generating function of the Legendre polynomial is given as
\begin{equation*}
\sum _{n=0}^{\infty } P_{n}( y) z^{n} =\frac{1}{\sqrt{1-2yz+z^{2}}}.
\end{equation*}
Thus, we have
\begin{align*}
\sum _{n=0}^{\infty } x^{n} P_{n}( y) z^{2n} & =\sum _{n=0}^{\infty } P_{n}( y)\left( xz^{2}\right)^{n}=\frac{1}{\sqrt{1-2xyz^{2} +x^{2} z^{4}}}
\end{align*}
and
\begin{align*}
\sum _{n=1}^{\infty } x^{n} P_{n-1}( y) z^{2n} & =xz^{2}\sum _{n=1}^{\infty } x^{n-1} P_{n-1}( y) z^{2( n-1)}=\frac{xz^{2}}{\sqrt{1-2xyz^{2} +x^{2} z^{4}}}
.\end{align*}
By using these relations, we get our statement for $\Delta_-\neq0$.
\end{proof}
\end{Proposition}

\section{Conclusion and Discussion}
In this study, we analyzed the return probability and its generating function of quantum and correlated random walks in the one-dimensional integer lattice for general settings. We proved that the return probability could be written in terms of the Legendre polynomial. In particular, the return probability of the quantum walk depends only on the absolute value of the first element, $|\alpha|$, of the coin matrix and does not depend on the initial state. Also, the return probability is independent of the initial state for the correlated random walk with $a = d$. Furthermore, we showed that the generating function of the return probability is expressed in terms of elliptic integrals of both the first and second kinds for the quantum walk. Our result generalizes the previous research \cite{konno2010quantum}.

Historically, comparisons between the quantum and classical walks have led to new insights and extended the potential of new theories and applications. We hope that this research will provide a mathematical foundation for the properties of quantum walks. For future research, a further analysis using the generating function of the return probability obtained in this study would be interesting. Moreover, extending the one-dimensional lattice to a higher dimensional lattice would be one of the fascinating problems.

\label{sec:summary}

\appendix
\section{Appendix}

\subsection{Proof of Proposition \ref{prop:qw_return}}
\label{appendix1}
First, \ from Lemma \ref{lemma:path}, we have
\begin{align*}
\Xi _{2n}^{( QW)} (n,n)\varphi  & =e^{2ni\theta }| \alpha | ^{2n}\sum _{\gamma =1}^{n}\left( -\frac{| \beta | ^{2}}{| \alpha | ^{2}}\right)^{\gamma }\left(\begin{array}{ l }
n-1\\
\gamma -1
\end{array}\right)^{2}\left[\begin{array}{ c c }
\frac{n}{\gamma } & \frac{n-\gamma }{\alpha \gamma } \beta -\frac{\overline{\alpha }}{\overline{\beta }}\\
-\frac{n-\gamma }{\overline{\alpha } \gamma }\overline{\beta } +\frac{\alpha }{\beta } & \frac{n}{\gamma }
\end{array}\right]\left[\begin{array}{ c }
\varphi _{1}\\
\varphi _{2}
\end{array}\right]\\
 & =e^{2ni\theta } |\alpha |^{2n}\sum _{\gamma =1}^{n}\left( -\frac{|\beta |^{2}}{|\alpha |^{2}}\right)^{\gamma }\left(\begin{array}{ c }
n-1\\
\gamma -1
\end{array}\right)^{2}\left[\begin{array}{ c }
\frac{n}{\gamma }\left(\frac{\alpha \varphi _{1} +\beta \varphi _{2}}{\alpha }\right) -\frac{\varphi _{2}}{\alpha \overline{\beta }}\\
\frac{\varphi _{1}}{\overline{\alpha } \beta } +\frac{n}{\gamma }\left(\frac{\overline{\alpha } \varphi _{2} -\overline{\beta } \varphi _{1}}{\overline{\alpha }}\right)
\end{array}\right] .
\end{align*}
Here,
\begin{align*}
 & \left| \sum _{\gamma =1}^{n}\left( -\frac{|\beta |^{2}}{|\alpha |^{2}}\right)^{\gamma }\left(\begin{array}{ l }
n-1\\
\gamma -1
\end{array}\right)^{2}\left(\frac{n}{\gamma }\left(\frac{\alpha \varphi _{1} +\beta \varphi _{2}}{\alpha }\right) -\frac{\varphi _{2}}{\alpha \overline{\beta }}\right)\right| ^{2}\\
 & =n^{2}\frac{| \alpha \varphi _{1} +\beta \varphi _{2}| ^{2}}{|\alpha |^{2}}\left\{\sum _{\gamma =1}^{n}\frac{1}{\gamma }\left( -\frac{|\beta |^{2}}{|\alpha |^{2}}\right)^{\gamma }\left(\begin{array}{ c }
n-1\\
\gamma -1
\end{array}\right)^{2}\right\}^{2}\\
 & \ \ \ \ \ +\frac{| \varphi _{2}| ^{2}}{|\alpha |^{2} |\beta |^{2}}\left\{\sum _{\gamma =1}^{n}\left( -\frac{|\beta |^{2}}{|\alpha |^{2}}\right)^{\gamma }\left(\begin{array}{ c }
n-1\\
\gamma -1
\end{array}\right)^{2}\right\}^{2}\\
 & \ \ \ \ \ -2n\Re \left(\frac{\varphi _{1}\overline{\varphi _{2}}}{\overline{\alpha } \beta } +\frac{| \varphi _{2}| ^{2}}{|\alpha |^{2}}\right)\sum _{\gamma =1}^{n}\frac{1}{\gamma }\left( -\frac{|\beta |^{2}}{|\alpha |^{2}}\right)^{\gamma }\left(\begin{array}{ c }
n-1\\
\gamma -1
\end{array}\right)^{2}\sum _{\delta =1}^{n}\left( -\frac{|\beta |^{2}}{|\alpha |^{2}}\right)^{\delta }\left(\begin{array}{ c }
n-1\\
\delta -1
\end{array}\right)^{2}
\end{align*}
and
\begin{align*}
 & \left| \sum _{\gamma =1}^{n}\left( -\frac{|\beta |^{2}}{|\alpha |^{2}}\right)^{\gamma }\left(\begin{array}{ c }
n-1\\
\gamma -1
\end{array}\right)^{2}\left\{\frac{\varphi _{1}}{\overline{\alpha } \beta } +\frac{n}{\gamma }\left(\frac{\overline{\alpha } \varphi _{2} -\overline{\beta } \varphi _{1}}{\overline{\alpha }}\right)\right\}\right| ^{2}\\
= & \frac{| \varphi _{1}| ^{2}}{|\alpha |^{2} |\beta |^{2}}\left\{\sum _{\gamma =1}^{n}\left( -\frac{|\beta |^{2}}{|\alpha |^{2}}\right)^{\gamma }\left(\begin{array}{ c }
n-1\\
\gamma -1
\end{array}\right)^{2}\right\}^{2}\\
 & \ \ +n^{2}\frac{| \overline{\alpha } \varphi _{2} -\overline{\beta } \varphi _{1}| ^{2}}{|\alpha |^{2}}\left\{\sum _{\gamma =1}^{n}\frac{1}{\gamma }\left( -\frac{|\beta |^{2}}{|\alpha |^{2}}\right)^{\gamma }\left(\begin{array}{ c }
n-1\\
\gamma -1
\end{array}\right)^{2}\right\}^{2}\\
 & \ \ +2n\Re \left(\frac{\varphi _{1}\overline{\varphi _{2}}}{\overline{\alpha } \beta } -\frac{| \varphi _{1}| ^{2}}{|\alpha |^{2}}\right)\sum _{\gamma =1}^{n}\frac{1}{\gamma }\left( -\frac{|\beta |^{2}}{|\alpha |^{2}}\right)^{\gamma }\left(\begin{array}{ c }
n-1\\
\gamma -1
\end{array}\right)^{2}\sum _{\delta =1}^{n}\left( -\frac{|\beta |^{2}}{|\alpha |^{2}}\right)^{\delta }\left(\begin{array}{ c }
n-1\\
\delta -1
\end{array}\right)^{2} ,
\end{align*}
where $\Re ( z)$ denotes the real part of a complex number $z$. Therefore, the return probability becomes
\begin{align*}
r_{2n}^{( QW)} (0) & =|\alpha |^{4n-2}\left[ n^{2}\left\{\sum _{\gamma =1}^{n}\frac{1}{\gamma }\left( -\frac{|\beta |^{2}}{|\alpha |^{2}}\right)^{\gamma }\left(\begin{array}{ c }
n-1\\
\gamma -1
\end{array}\right)^{2}\right\}^{2}\right. \\
 & \ \ \ \ \ \ \ +\frac{1}{|\beta |^{2}}\left\{\sum _{\gamma =1}^{n}\left( -\frac{|\beta |^{2}}{|\alpha |^{2}}\right)^{\gamma }\left(\begin{array}{ c }
n-1\\
\gamma -1
\end{array}\right)^{2}\right\}^{2}\\
 & \ \ \ \ \ \ \left. -2n\sum _{\gamma =1}^{n}\frac{1}{\gamma }\left( -\frac{|\beta |^{2}}{|\alpha |^{2}}\right)^{\gamma }\left(\begin{array}{ c }
n-1\\
\gamma -1
\end{array}\right)^{2}\sum _{\delta =1}^{n}\left( -\frac{|\beta |^{2}}{|\alpha |^{2}}\right)^{\delta }\left(\begin{array}{ l }
n-1\\
\delta -1
\end{array}\right)^{2}\right] .
\end{align*}
Moreover, we will rewrite $\Xi _{2n}^{( QW)} (n,n)$ by using the Jacobi polynomial, $P_{n}^{( \nu ,\mu )} (x)$, which is orthogonal on $[-1,1]$ with respect to $(1-x)^{\nu } (1+x)^{\mu }$ for $\nu ,\mu  >-1$. The following relation holds:
\begin{align*}
P_{n}^{( \nu ,\mu )}( x) & =\frac{\Gamma ( n+\nu +1)}{\Gamma ( n+1) \Gamma ( \nu +1)} \ _{2} F_{1}\left( -n,n+\nu +\mu +1;\nu +1;\frac{1-x}{2}\right) ,
\end{align*}
where $\Gamma (z)$ is the gamma function and $_{2} F_{1}( a,b;c;z)$ is the hypergeometric function, which satisfies \ 
\begin{equation*}
_{2} F_{1} (a,b;c;z)=(1-z)^{-a}{}_{2} F_{1}\left( a,c-b;c;\frac{z}{z-1}\right) .
\end{equation*}
By putting $k=2|\alpha |^{2} -1$ and using these relations, we can write
\begin{align}
\sum _{\gamma =1}^{n}\frac{1}{\gamma }\left( -\frac{| \beta | ^{2}}{| \alpha | ^{2}}\right)^{\gamma }\left(\begin{array}{ c }
n-1\\
\gamma -1
\end{array}\right)^{2} & =-\frac{| \beta | ^{2}}{| \alpha | ^{2}} \ _{2} F_{1}\left( 1-n,1-n;2;-\frac{| \beta | ^{2}}{| \alpha | ^{2}}\right)\nonumber \\
 & =-| \beta | ^{2}\left(\frac{1}{| \alpha | ^{2}}\right)^{n}{}_{2} F_{1}\left( -(n-1),n+1;2;| \beta | ^{2}\right)\nonumber\\
 \label{eq:geo1}
 & =-\frac{| \beta | ^{2}}{n}\left(\frac{1}{| \alpha | ^{2}}\right)^{n} P_{n-1}^{(1,0)} (k).
\end{align}
Similarly,
\begin{align}
\label{eq:geo2}
\sum _{\gamma =1}^{n}\left( -\frac{| \beta | ^{2}}{| \alpha | ^{2}}\right)^{\gamma }\left(\begin{array}{ c }
n-1\\
\gamma -1
\end{array}\right)^{2} & =-| \beta | ^{2}\left(\frac{1}{| \alpha | ^{2}}\right)^{n} P_{n-1}^{(0,0)} (k).
\end{align}
Thus, we can rewrite $r_{2n}^{( QW)}( 0)$ with the Jacobi polynomial.
\begin{equation*}
r_{2n}^{( QW)} (0)=\frac{|\beta |^{2}}{|\alpha |^{2}}\left[ |\beta |^{2}\left\{P_{n-1}^{(1,0)} (k)\right\}^{2} +\left\{P_{n-1}^{(0,0)} (k)\right\}^{2} -2|\beta |^{2} P_{n-1}^{(1,0)} (k)P_{n-1}^{(0,0)} (k)\right] .\ 
\end{equation*}
Also, using the relation from \cite{andrews1999special}, 
\begin{align*}
( n+\nu +1) P_{n}^{( \nu ,\mu )}( x) -( n+1) P_{n+1}^{( \nu ,\mu )}( x) & =\frac{( 2n+\nu +\mu +2)( 1-x)}{2} P_{n}^{( \nu +1,\mu )}( x) ,
\end{align*}
we have the following by setting $\nu =\mu =0$ and $x=k$:
\begin{align}
\label{eq:jacobi}
P_{n-1}^{( 1,0)}( k) =\frac{P_{n-1}^{( 0,0)}( k) -P_{n}^{( 0,0)}( k)}{2| \beta | ^{2}} .
\end{align}
Therefore, we get 
\begin{equation*}
r_{2n}^{( QW)}( 0) =\frac{\left\{P_{n-1}^{(0,0)} (k)\right\}^{2} -2kP_{n}^{(0,0)} (k)P_{n-1}^{(0,0)} (k)+\left\{P_{n}^{(0,0)} (k)\right\}^{2}}{2( k+1)} .\ 
\end{equation*}
Since $P_{n}^{(0,0)} (x)=P_{n} (x)$ holds, the proof is complete.
\qed
\subsection{Proof of Proposition \ref{prop:crw_return}}
\label{appendix2}
Lemma \ref{lemma:path} can also be applied to correlated random walks. Thus we have
\begin{align*}
\Xi _{2n}^{( CRW)} (n,n)\hat\varphi  & =(ad)^{n}\sum _{\gamma =1}^{n}\left(\frac{bc}{ad}\right)^{\gamma }{\displaystyle \binom{n-1}{\gamma -1}}^{2}\begin{bmatrix}
\frac{n}{\gamma } & \frac{b}{a}\frac{n-\gamma }{\gamma } +\frac{d}{c}\\
\frac{c}{d}\frac{n-\gamma }{\gamma } +\frac{a}{b} & \frac{n}{\gamma }
\end{bmatrix}\begin{bmatrix}
\hat{\varphi }_{1}\\
\hat{\varphi }_{2}
\end{bmatrix}\\
 & =(ad)^{n}\sum _{\gamma =1}^{n}\left(\frac{bc}{ad}\right)^{\gamma }{\displaystyle \binom{n-1}{\gamma -1}}^{2}\begin{bmatrix}
\frac{n}{\gamma }\hat{\varphi }_{1} +\left(\frac{b}{a}\frac{n-\gamma }{\gamma } +\frac{d}{c}\right)\hat{\varphi }_{2}\\
\left(\frac{c}{d}\frac{n-\gamma }{\gamma } +\frac{a}{b}\right)\hat{\varphi }_{1} +\frac{n}{\gamma }\hat{\varphi }_{2}
\end{bmatrix} .
\end{align*}
The return probability of the correlated random walk becomes
\begin{align*}
r_{2n}^{( CRW)}( 0) & =\| \Xi _{2n}^{( CRW)} (n,n)\hat\varphi \| _{1}\\
 & =( ad)^{n}\sum _{\gamma =1}^{n}\left(\frac{bc}{ad}\right)^{\gamma }\left(\begin{array}{ l }
n-1\\
\gamma -1
\end{array}\right)^{2}\\
 & \ \ \ \ \times \left\{\frac{n}{\gamma }\left(\frac{ac\hat{\varphi }_{1} +bd\hat{\varphi }_{2}}{ad} +1\right) +\frac{ad-bc}{abcd}( ac\hat{\varphi }_{1} +bd\hat{\varphi }_{2})\right\} .
\end{align*}
When $\Delta_-\neq0$, it follows from relations (\ref{eq:geo1}) and (\ref{eq:geo2}) that
\begin{align*}
\sum _{\gamma =1}^{n}\frac{1}{\gamma }\left(\frac{bc}{ad}\right)^{\gamma }\left(\begin{array}{ c }
n-1\\
\gamma -1
\end{array}\right)^{2} & =\frac{bc}{nad}\left(\frac{\Delta _{-}}{ad}\right)^{n-1} P_{n}^{( 1,0)}\left(\frac{\Delta _{+}}{\Delta _{-}}\right)
\end{align*}
and
\begin{align*}
\sum _{\gamma =1}^{n}\left(\frac{bc}{ad}\right)^{\gamma }\left(\begin{array}{ c }
n-1\\
\gamma -1
\end{array}\right)^{2} & =\frac{bc}{ad}\left(\frac{\Delta _{-}}{ad}\right)^{n-1} P_{n}^{( 0,0)}\left(\frac{\Delta _{+}}{\Delta _{-}}\right) .
\end{align*}
Therefore,
\begin{align*}
 & r_{2n}^{( CRW)}( 0) =\frac{\Delta _{-}^{n-1}}{ad}\left( \Delta _{-}( ac\hat{\varphi }_{1} +bd\hat{\varphi }_{2}) P_{n-1}^{( 0,0)}\left(\frac{\Delta _{+}}{\Delta _{-}}\right) +bck_{+} P_{n-1}^{( 1,0)}\left(\frac{\Delta _{+}}{\Delta _{-}}\right)\right) .
\end{align*}
By (\ref{eq:jacobi}), this can be converted to
\begin{equation*}
\frac{\Delta _{-}^{n}}{2ad}\left( k_{-} P_{n-1}^{( 0,0)}\left(\frac{\Delta _{+}}{\Delta _{-}}\right) +k_{+} P_{n}^{( 0,0)}\left(\frac{\Delta _{+}}{\Delta _{-}}\right)\right) .
\end{equation*}
Replacing $P_{n}^{(0,0)} (x)$ with $P_{n} (x)$, we get the first expression in the statement. When $\Delta_-= 0$ and $ a=b=p,c=d=q=1-p$, we have
\begin{align*}
r_{2n}^{( RW)}( 0) & =2n( pq)^{n}\sum _{\gamma =1}^{n}\frac{1}{\gamma }\left(\begin{array}{ l }
n-1\\
\gamma -1
\end{array}\right)^{2}.\end{align*}
Using the following relation, we obtain the desired conclusion.
\begin{align*}
2n\sum _{\gamma =1}^{n}\frac{1}{\gamma }\left(\begin{array}{ l }
n-1\\
\gamma -1
\end{array}\right)^{2} & =\binom{2n}{n}.
\end{align*}
\qed
\printbibliography
\end{document}